\documentclass[12pt, preprint]{aastex}

\shorttitle{P Cygni Analog in M31}
\shortauthors{Massey}

\begin{document}

\title{The Discovery of a P Cygni Analog in M31\altaffilmark{1}}

\author{Philip Massey\altaffilmark{2}}

\affil{Lowell Observatory, 1400 W. Mars Hill Road, Flagstaff, AZ 86001}
\email{Phil.Massey@lowell.edu}

\altaffiltext{1}{Based in part on observations made with the NASA/ESA
Hubble Space Telescope, obtained from the Data Archive
at the Space Telescope Science
Institute, which is operated by the Association for Universities for
Research in Astronomy (AURA), Inc., under NASA contract NAS 5-26555.}
\altaffiltext{2} {Visiting Astronomer, Kitt Peak National Optical 
Astronomy Observatory, which is operated by AURA, Inc., under cooperative agreement
with the National Science Foundation (NSF).}

\begin{abstract}

We present spectroscopy and discuss the  photometric history of a previously obscure
star in M31.  The spectrum of the star is an extremely close match to that of P~Cygni, one of
the archetypes of  Luminous Blue Variables (LBVs).
The star has not shown much variability over the past 40 years ($<0.2$~mag), although 
small-scale (0.05~mag) variations over a year appear to be real.
Nevertheless, the presence of a sub-arcsecond extension around the star is
indicative of a past outburst, and from the nebula's size (0.5~pc diameter) we estimate the
outburst took place
roughly 2000 yrs ago.  P Cygni itself exhibits a similar photometric behavior, and has
a similar nebula (0.2~pc diameter).  We argue that this may be more typical behavior
for LBVs than commonly assumed.  The star's location in the HR diagram 
offers substantial support for stellar evolutionary models that include the effects
of rotation, as the star is just at a juncture in the evolutionary track of
a 85$M_\odot$ star.  The star is likely
in a transition from an O star to a late-type WN Wolf-Rayet.  
\end{abstract}


\section{Introduction}
\label{Sec-intro}

Luminous Blue Variables (LBVs) are a rare class of luminous stars which
undergo episodic mass-loss, and probably represent a transition between 
the most massive O stars and the red supergiant and/or Wolf-Rayet stage.
(See the recent conference proceedings edited by Davidson et al.\ 1988,
Nota \& Lamers 1997, and de Groot \& Sterken 2001.)
All of the ``accepted" LBVs were discovered
on the basis photometric and/or spectroscopic variability, but the number of
LBV ``candidates" (based upon spectroscopic similarities to known LBVs)
exceeds this number (Parker 1997).  Understanding the number of bona-fide
LBVs relative to that of other evolved massive stars (Wolf-Rayets, 
red supergiants) in nearby
galaxies (i.e., at differing metallicities) provides a crucial
test of the success of modern stellar evolutionary theory (Maeder \& Meynet
2000).  Massey (2002) has argued that if the archetypal LBVs P~Cygni or $\eta$ Carina
were located in M~31 or M~33 we would not know of them today, since their spectacular
photometric outbursts were hundreds of years ago.  

In this {\it Letter} we report the discovery of a star in M31 whose spectrum is uniquely
similar to that of P Cygni itself.  A comparison of the star's location with stellar
evolutionary tracks provides a strong substantiation of the latest generation of stellar
models that include the effects of rotation (Meynet \& Maeder 2005),
while also offering a caution in interpreting evolutionary stages of these hydrostatic
models.  The star has been relatively photometrically stable for the past 40 years,
yet it shows compelling evidence of an outburst 2 millennia ago.  This provides 
further support for the argument large photometric variations on the time-scale of
decades may be more a selection effect than a consequence of the physics of LBVs.

\section{The Discovery}

The Local Group Galaxies Survey (LGGS) project has obtained 
images of M31 with the Kitt Peak 4-m telescope.
The LGGS has now completed the photometric calibrations and photometry
for M31 ($>$350,000 stars), and is preparing the paper for publication.
A subset of several dozen
bright, blue stars were selected for spectroscopy with the WIYN\footnote{The WIYN Observatory is a joint facility of the University
of Wisconsin-Madison, Indiana University, Yale University, and NOAO.} 3.5-m telescope and Hydra multi-object fiber positioner.  
The selection criteria included $V<18$, $B-V<0.5$, and the reddening free index $Q=(U-B)-0.72(B-V)<-0.6$ in order to eliminate foreground dwarfs.  The spectra were obtained using a 790 l mm$^{-1}$
grating (KPC-18C) in second order, with a BG-39 blocking filter to filter out first order red light.
The observations covered 3970\AA\ to 5030\AA.  The ``blue" fiber bundle was used for these
observations; each fiber is 3.1" in diameter.  The detector was a Tektronix 2048 x 2048 device
(``T2KA") with 24$\mu$m pixels.  The resulting spectral resolution was 1.5\AA.
Our exposure of the northern field (in which our object was found) consisted of six 30m integrations
(i.e., 3 hrs).  The field was reconfigured halfway through the sequence in order to update the
apparent position changes due to refraction.  The observation was made on (UT) 2005 September 29.

\begin{figure}
\plotone{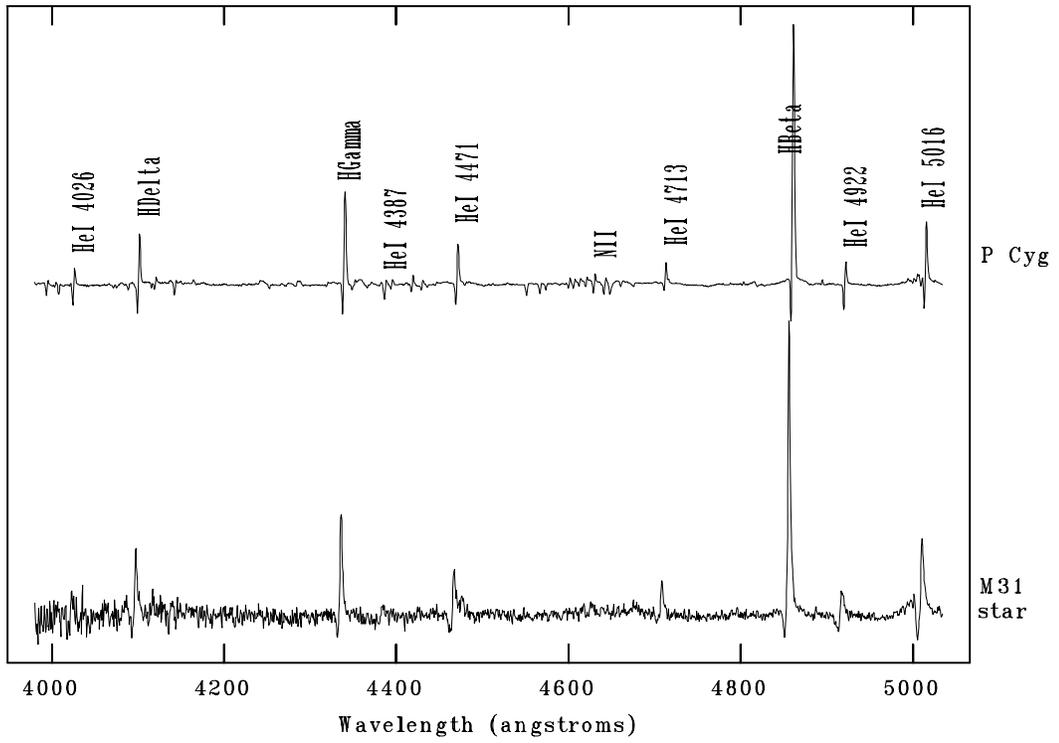}
\caption{\label{fig:spectra} Comparison of the spectrum of LGGS J004341.84+411112.0 with
P~Cygni.  Major lines have been identified; see also Stahl et al.\ (1993)}
\end{figure}

In reducing the data that night we were immediately struck by the similarity of one of the stars to
the well-known Galactic star P Cygni.   The following night we obtained a 10 s exposure
of P Cygni
itself with the same setup to confirm our memory.   We give the comparison in Fig.~\ref{fig:spectra}.

The degree of similarity to P~Cygni is nearly unique.
While the term ``P Cygni profile"  (blue-shifted absorption with an emission component extending redwards  from line center) is used to describe individual lines in other stars having this characteristic
mass-loss signature, only a few stars show even
a close resemblance to P Cygni itself throughout the blue-optical region, 
and all of these stars have been called LBVs.  Even
amongst these the differences can be striking.  For instance,  although the LMC LBV-candidate R81 has been described as the ``P Cygni of the LMC" by Wolf et al.\ (1981a), its optical
spectrum contains  He~I lines without a strong emission component; compare their Fig.~1 with
our Fig.~\ref{fig:spectra}.
Walborn (1989)  describes other LBVs at minimum light (i.e., maximum $T_{\rm eff}$) as having
the same general spectral characteristics as P Cygni, but the comparison is more apt in
some cases than in others.
AG~Car (Hutsem\'{e}kers \& Kohoutek 1988, Fig.~1), 
AF And, M33 Var B (Kenyon \& Gallagher 1985, Figs.~2-3), and R127 (Stahl et al.\ 1983, Fig.~3)
do all show the same general P Cygni lines in He~I and in the Balmer lines as does P Cygni, 
but R71 is an another example where the He~I lines do not show a pronounced 
emission component (Wolf et al.\ 1981b, Fig.~3).  HDE 316285 also shows many
similarities in the infrared (McGregor et al.\ 1988) and yellow-red (Morrison \& Rao (1990),
but the blue-optical spectra shows little resemblance to P Cygni (Walborn \& Fitzpatrick 2000).
In the Milky Way, the LBV He 3-519 (Smith et al.\ 1994) probably bears the most similar spectrum
known to P Cygni.

\begin{figure}
\epsscale{0.72}
\plotone{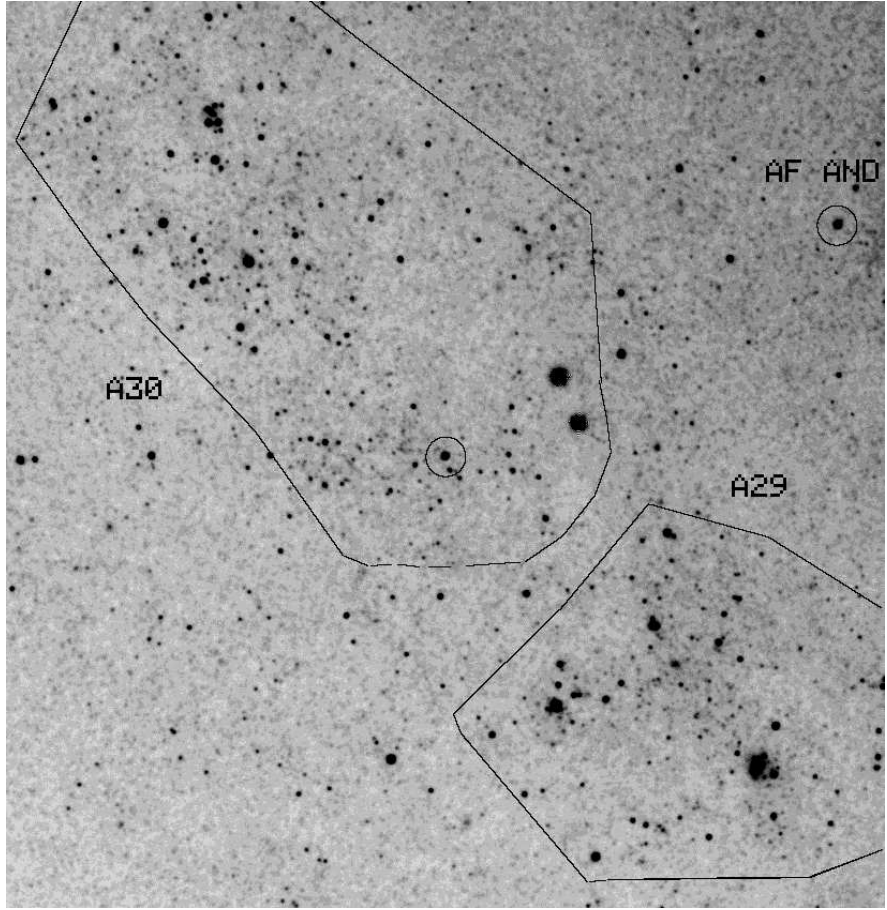}
\caption{\label{fig:fc} The location of LGGS J004341.84+41112.0. This star circled in the middle
of this $V$  image (from the LGGS)
is the newly found P Cygni analog.  The well-known LBV AF And is located 
2' to the NW.  The circles have a diameter of 10".
The outlines of the  two OB associations A30 and A29 (van den Bergh 1964),
based upon Hodge (1981).   
}
\end{figure}

Our M31 star is identified in the LGGS catalog as J004341.84+411112.0, 
based upon its 
J2000 coordinates
(i.e., $\alpha=00^{\rm h}43^{\rm m}41.\!^{\rm s}84$, $\delta=+41^\circ11' 12.\!"0$) determined from
USNO-B astrometric positions in the field.  It is located in A30, an OB
association cataloged by van den Bergh (1964); see Fig.~\ref{fig:fc}.
We note with some amusement that one of the original five Hubble-Sandage variables (Hubble \& Sandage 1953), Var 19 = AF And,
is located just 2' away\footnote{We eliminated the possibility that we mistakenly observed
AF And itself.  First, the two stars are clearly distinct.  Brian Skiff
kindly compared our field to the finding chart of AF And (Variable 19) from Hubble (1929), the  paper
which first showed that M31 is a galaxy(!), and confirmed the J2000 coordinates from the Two Micron All
Sky Survey (2Mass) as $\alpha=00^{\rm h}43^{\rm m}33.\!^{\rm s}08$, $\delta=+41^\circ12'10.\!"3$ (uncertainty $<0.\!"2$), which led to a revision by SIMBAD in their listed 
position by a few arcseconds.  Second, we considered whether or not it was possible
for the fiber positioner to have accidently {\it dropped} the fiber button, miraculously hitting the
position of AF And rather than our intended target.  It was not.  For one thing, we made a minor tweak in
the positions of all of the fibers half way through the exposure (see above), and the
spectrum before
and after the reconfiguration matched.  In addition, had the button been dropped the positioner would
have been unable to retrieve it when we reconfigured for the next field, and that was not the
case.  We are indebted to Nolan Walborn for raising this particular specter, but we feel it has
been completely exorcised.}.

The star was first cataloged (as ``40 4631") by 
Berkhuijsen et al.\ (1988), who presented {\it UBVR$_J$} photometry based upon photographic plates
taken in 1963.  The next published photometry is given by Magnier et al.\ (1992) from their  1990
{\it BVRI} CCD survey of a 1 deg$^2$ field in M31, where it appears as object 228465.
 If the data are taken at face value, the star  brightened by 0.6 mag in $V$, and reddened by 0.5~mag in $B-V$ during
 the 37 years  (see Table~\ref{tab:phot}).  However,  nearby ($<10'$) stars of a similar
 brightness
 show this same offset in $V$, so we discount this as an
 indication of variability.  However, the photographic $B$ value of neighboring stars
 are more in accord with Magnier et al.\ (1992), albeit with large scatter.
 The LGGS CCD photometry, carried out in 2001-2002, agrees well with Magnier et al.\ (1992).  The slight differences
 between our two observations (0.06 mag) appears to be real: the formal errors are one-tenth of that, and a comparison
 of the stars in overlap between the two fields indicate median differences $<0.01$ at all filters except
 $I$, for which
 there is  a small systematic offset  (0.02~mag) for the 2002 observation.  G. Jacoby also
kindly obtained current-epoch (2005 November) 
$V$ and $R$  images at WIYN for us, which we have calibrated using the LGGS 
photometry of stars of similar color.  We conclude
 that over recent times the star has been relatively constant ($<$0.2~mag) in brightness over the
 past 40 years, but that it does show real variations of order 0.05~mag.  This behavior is quite
 similar to that of  P Cygni, although different than many other LBVs (Israelian \& de Groot 1999).
 
 Hill et al.\ (1995) identify their UV-bright source 30-3 with this star, using  Magnier et al.\ (1992)
 as a cross-reference.
 They used {\it Ultraviolet Imaging Telescope (UIT)} data, and
 detected the star in the 2500\AA\ image but not in the 1500\AA\ image\footnote{They also had ground-based {\it B}  and {\it R}
 CCD photometry indicating a
 magnitude of $B=18.29$, which is about 0.3~mag fainter at $B$ than what was seen by any of the other studies; they
 estimate their errors as 0.05-0.10 mag, but provide no details (date or telescope) of the
 observations.
 One can infer that the observation was likely made in the early 1990s, so similar in time to Magnier et al.\ (1992).
 They report $R=17.12$, which is in good agreement with the  values found by Magnier et al.\ (1992) and the LGGS.}. The fact that the star is UV-bright is consistent with the Massey et al.\ (1996)
 study of {\it UIT} sources in M33, which discovered many LBV candidates.
  
\begin{figure}
\epsscale{0.72}
\plotone{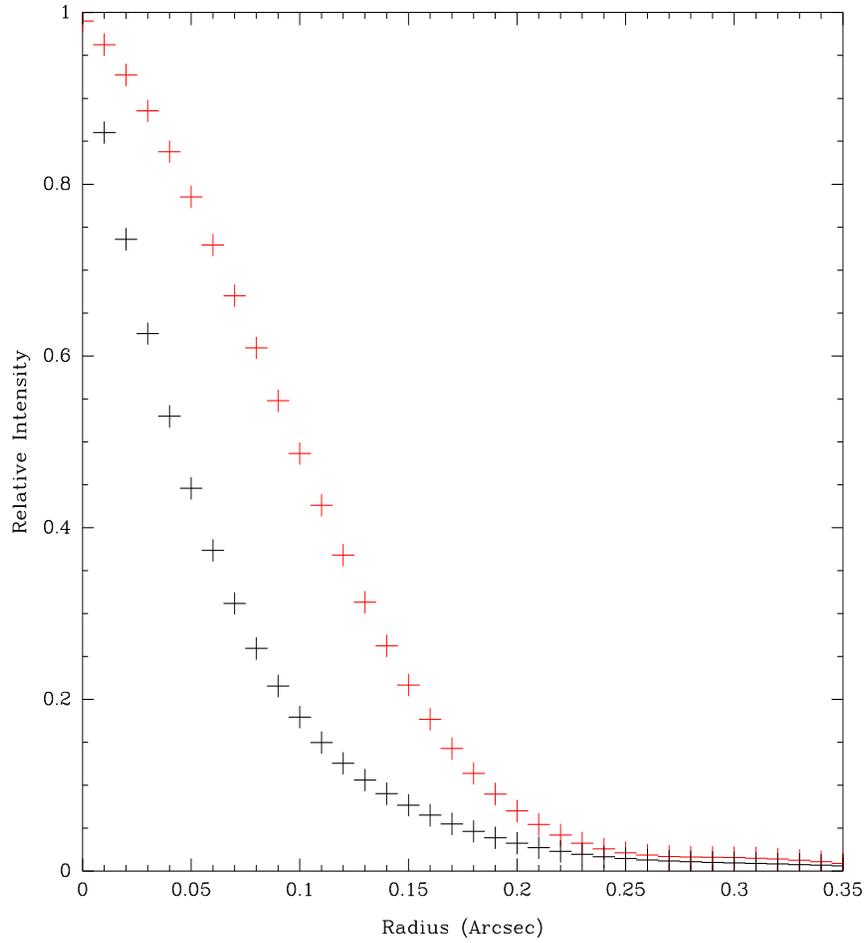}
\caption{\label{fig:radplot} Spatial extension of LGGS J004341.84+411112.0.  The radial profile of
J004341.84+4111112.0 was measured from a WFPC2 image (WF3 chip), and is shown in red.
The profile of another star on the same WF3 image is shown in black for comparison.
}
\end{figure}

We checked the {\it HST} data archives for images of the star; only one WFPC2 image
contains a (barely) unsaturated exposure.   This is a 300 s exposure through the F675W (``R")
filter, u42z5701r.  The star is in fact slightly extended, with a full-width-at-half-maximum
(fwhm) of 2.0 pixels (0.2") of the WF3 chip.  Other stars on the image have significantly
smaller fwhms, as shown in Fig.~\ref{fig:radplot}.
However, the star does not appear to be double, given
the radial profile, just extended.  If
we make the simple assumption that the size of the image
adds in quadrature, then there is something around the star that has a diameter of about
$0.\!"015$.  At the distance of M31 (760 kpc, according to van den Bergh 2000), this would
have a size of 0.5~pc.     P Cygni is known to have several shells of gas around it; the most
prominent of these has a diameter of 22" (0.2~pc), which is taken as evidence that the star had a major
mass-loss event 900 years ago (see discussion in Israelian \& de Groot 1999).
 If the expansion velocity is the same for our object, then
the relevant time scale would be about 2,000 yr.  Confirmation with the better scale of the ACS
would be desirable; several ACS images in the archive include the field, but the star is 
saturated\footnote{Our own scheduled parallel imaging of this region did not happen
as a result of the failure of the Space Telescope Imaging Spectrograph.}.

\begin{figure}
\epsscale{0.72}
\plotone{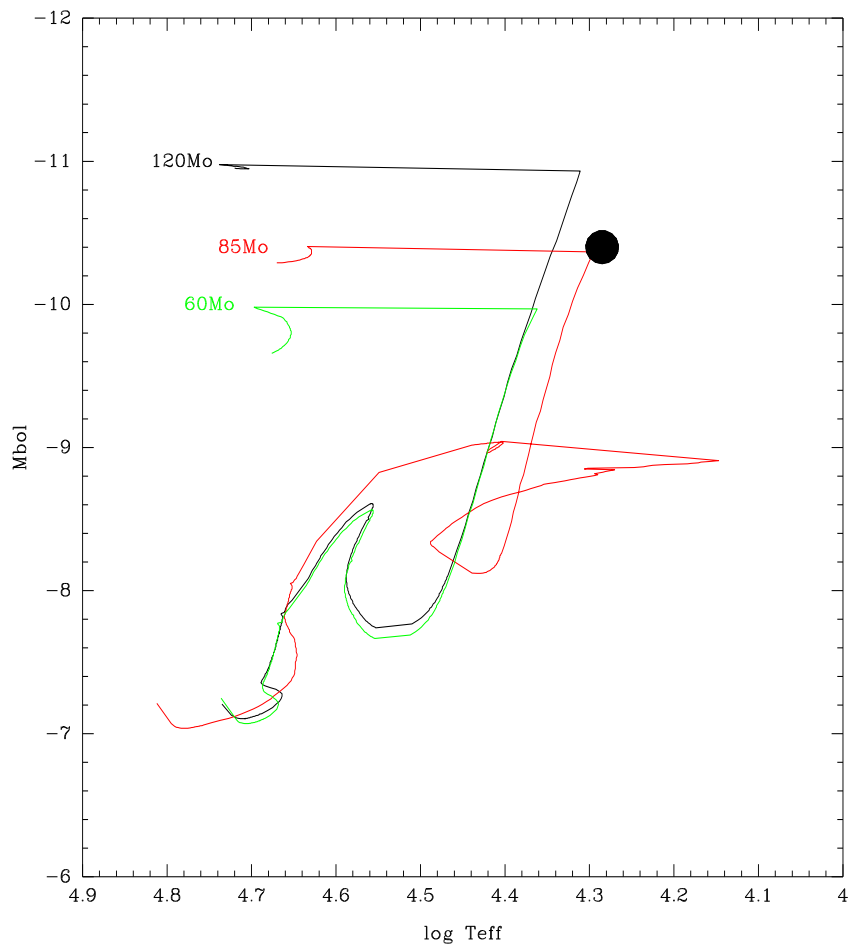}
\caption{\label{fig:tracks}.  The location of M31's P Cygni analog in the H-R diagram.
We have plotted the location of LGGS J004341.84+411112.0 as a black dot, where we
have {\it assumed} that its  
bolometric correction and effective temperatures are  the same as those of P Cygni (Lamers
et al.\ 1983).  The evolutionary tracks of Meynet \& Maeder (2005) for an initial rotation
velocity of 300 km s$^{-1}$ are shown as solid colored lines,
with the initial masses indicated.  The star would appear to be at the coolest point  in the
evolution of a 85$M_\odot$ star, likely a transition point between an O star and a Wolf-Rayet.
}
\end{figure}

Determining the star's absolute visual magnitude requires knowing the reddening.  If 
the star's intrinsic color is the same as P~Cygni's, $(B-V)_0=-0.19$, (Lamers et al.\ 1983),
then $E(B-V)=0.64$.   The foreground reddening to M31 is $E(B-V)=0.07$, 
but most of the
reddening of OB stars in M31 is due to extinction within the disk of M31 (van den Bergh 2000).  
Four OB associations
away from the minor axis were found to have $E(B-V)=0.12-0.24$ by Massey et al.\ (1986).
The reddening across M31
is quite patchy, and A30's location is near the minor axis where inclination effects will
result in a larger amount of dust through the line of sight, so the value $E(B-V)=0.64$ seems
quite reasonable.
We derive an absolute visual magnitude of $M_V=-8.9$. 
This may be compared to 
P Cygni's absolute visual magnitude, $M=-8.3$  (Lamers et al.\  1983).   If the effective temperatures 
(and hence bolometric corrections) are similar as well, as is
implied by the strong spectral similarities,  we can scale P Cygni's  bolometric luminosity by 0.6~mag
as well, and expect that our star has $L/L_\odot=10^{6.1}$.
In Fig.~\ref{fig:tracks} we show the location of the star in the H-R diagram along with the
evolutionary tracks for z=0.04 (corresponding to the metallicity of M31's HII regions according to Zaritsky et al.\ 1994) from Meynet \& Maeder (2005).
According to their calculations, an object in this part of the H-R diagram would be
spectroscopically identified as a late-type WN (WNL) Wolf-Rayet star,
and should never undergo an LBV phase as such.
However, Andre Maeder (private communication) emphasizes that since their calculations are
 hydrostatic one can not really tell what stage a star is in (LBV vs WNL) from the models,
as the difference between the
two may be primarily one of atmospheric extent.  The star may be at a transition point, and may
evolve to the WNL stage after episodes of mass loss.  There are hints in the spectra that the
star is nitrogen-rich (Fig.~\ref{fig:spectra}), as is P Cygni and many other LBVs
(Walborn 1989); a better spectrum is needed for modeling the CNO 
abundances, and such work is planned.

\section{Summary and Discussion}

The LGGS project has obtained photometry of $>$350,000 stars seen towards M31.  One of the
brightest of these, LGGS 004341.84+411112.0, turns out to have a spectrum that is remarkably
similar to that of the P Cygni, one of the best studied LBVs in the Milky Way.   
The star has been relatively constant
in $B$ over the past 40 years, with variations $<0.2$~mag, but with evidence of variations of 0.05~mag
during a year.  Much the same can be said for P Cygni (Israelian \& de Groot 1999).
However, the star has likely had an outburst two millennia ago, as judged by the fact that the
star is slightly extended {\it HST} images, indicative of a 0.5~pc nebula, similar to the 0.2~pc
nebula seen around P~Cygni.  This has consequences for interpreting the nature of objects
found with spectroscopic similarity to other known LBVs; these stars are considered LBV 
``candidates" until variability is established, but the present study emphasizes that this may
require more than a lifetime.

If we assume that the spectral similarity of our star to that of P Cygni is indicative of a similar
effective temperature (and hence intrinsic color and
bolometric correction), we can place the star on the H-R diagram
using our photometry.  The star falls on the evolutionary track for a 85$M_\odot$ star at the
extreme end its evolution to cooler temperature, just  as the evolutionary tracks turns to
lower luminosity (Meynet \& Maeder 2005).  The star is likely in a transition between an O star
and a WNL.  This provides strong vindication of the current generation of stellar evolutionary
models that include rotation, but it also notes the difficulties in interpreting the evolution
phases implied by the models.
An atmospheric analysis of the star is planned, and long-term follow-up (both photometric and
spectroscopic) is warranted.  

\acknowledgements

We are grateful for suggestions by an anonymous referee, as well as help from
George Jacoby and Brian Skiff.
Andre Maeder and
Georges Meynet made very useful comments on the issue of the evolutionary status of this
object.   The spectroscopic
observations were carried out with aid from Daryl Willmarth, Di Harmer, George Will, and
Hillary Mathis.  Parts of this project were funded by the NSF through grant 0093060,
and by NASA through a grant (GO-9794) from the Space Telescope Science Institute,
which is is operated by AURA, Inc., under NASA contract NAS 5-26555.

\clearpage

\begin{deluxetable}{l c c c c c c c c c c l}
\tabletypesize{\scriptsize}
\tablewidth{0pc}
\tablenum{1}
\tablecolumns{12}
\tablecaption{\label{tab:phot}Photometry of LGGS 004341.84+411112.0}
\tablehead{
\colhead{Date}  & 
\colhead{$V$}  & \colhead{$\sigma_V$} & 
\colhead{$B-V$} & \colhead{$\sigma_{B-V}$} & 
\colhead{$U-B$} & \colhead{$\sigma_{U-B}$} & 
\colhead{$V-R$\tablenotemark{a}} & \colhead{$\sigma_{V-R}$} &
\colhead{$R-I$\tablenotemark{a}} & \colhead{$\sigma_{R-I}$} &
\colhead{Ref/ID\tablenotemark{b}} 
}
\startdata
1963 Aug/Sep    & 18.13   & 0.19    &-0.14   & 0.29   & -0.31   & 0.26         & \nodata\tablenotemark{c}    & \nodata\tablenotemark{c}  &\nodata & \nodata  &[BHG88] 40 4631\\
1990 Sep 12-27 & 17.531 & 0.061 & 0.407 & 0.088 & \nodata & \nodata & 0.408 & 0.097 & 0.229 & 0.086 &[MLV92] 228465\\
2001 Sep 20-22 & 17.583 & 0.004 & 0.432 & 0.005 & -0.750 & 0.004      & 0.387 & 0.006 & 0.222  & 0.006 &LGGS 004341.84+411112.0\\
2002 Sep 11       & 17.511 & 0.005 & 0.478 & 0.005 & -0.771 & 0.005      & 0.421 & 0.005 & 0.332: & 0.005 & LGGS 004341.84+411112.0 \\
2005 Nov 7         & 17.458 & 0.005 & \nodata & \nodata & \nodata & \nodata & 0.411 & 0.005 & \nodata & \nodata & New \\
\enddata
\tablenotetext{a}{$R$ and $I$ are on the Kron-Cousins system.}
\tablenotetext{b}{BHG88=Berkhuijsen et al.\ 1988; MLV92=Magnier et al.\ 1992; LGGS=Massey
et al.\ in prep.}
\tablenotetext{c}{Berkhuijsen et al.\ 1988 do report {\it R} photometry, but it appears to be on the
Johnson rather than Kron-Cousins system.}
\end{deluxetable}

\clearpage

\clearpage

\clearpage

\clearpage

\end{document}